\documentclass[pre,twocolumn,showpacs,showkeys,preprintnumbers,amsmath,amssymb]{revtex4-1}
\usepackage{amscd,graphicx,color,colordvi,bm,amsfonts,stmaryrd}
\usepackage{dsfont}
\usepackage{hyperref}

\newcommand{\access}{\shortrightarrow}

\newcommand{\Prob}{P}
\newcommand{\Pn}[1]{P^{(#1)}}
\newcommand{\Tran}{P^{\phantom{j}}}

\newcommand{\Amat}{\mathds{A}}
\newcommand{\Bmat}{\mathds{B}}
\newcommand{\Dmat}{\mathds{D}}

\newcommand{\Id}{\mathds{1}}
\newcommand{\Zero}{\mathds{O}}
\newcommand{\Pmat}{\mathds{P}}
\newcommand{\Pmatn}[1]{\mathds{P}^{(#1)}}

\newcommand{\Mmat}{\mathds{M}}
\newcommand{\Qmat}{\mathds{Q}}
\newcommand{\Tmat}{\mathds{P}}
\newcommand{\Tmatperm}{\widetilde{\mathds{P}}}
\newcommand{\Tmattt}{\Tmatperm_{\Tset \shortrightarrow \Tset}}
\newcommand{\Tmattr}{\Tmatperm_{\Tset \shortrightarrow \Rset}}
\newcommand{\Tmatrr}{\Tmatperm_{\Rset \shortrightarrow \Rset}}

\newcommand{\Pex}{\Pmat^\text{exit}}
\newcommand{\Pexit}[2]{\Pex_{#1,#2}}

\newcommand{\Transient}{\mathds{T}}
\newcommand{\Trantorec}{\mathds{S}}
\newcommand{\Recurrent}{\mathds{R}}

\newcommand{\Tset}{\mathcal{T}}
\newcommand{\Rset}{\mathcal{R}}

\newcommand{\pivec}{\boldsymbol{\pi}}
\newcommand{\Cesaro}{\mathds{C}}
\newcommand{\evec}{\mathbf{e}}
\newcommand{\pvec}{\mathbf{p}}
\newcommand{\vvec}{\mathbf{v}}
\newcommand{\xvec}{\mathbf{x}}

\begin{document}


\title{%
  Random Walk on Lattices: Graph Theoretic Approach to Modeling Epitaxially Grown Thin Film
}

\author{Surachate Limkumnerd}
\email{surachate.l@chula.ac.th}
\affiliation{%
Department of Physics, Faculty of Science, Chulalongkorn University, Phayathai Rd., Patumwan, Bangkok 10330, Thailand \\
Research Center in Thin Film Physics, Thailand Center of Excellence in Physics, CHE, 328 Si Ayutthaya Rd., Bangkok 10400, Thailand
}

\date{\today}

\begin{abstract}
	Immense interests in thin-film fabrication for industrial applications have driven both theoretical and computational aspects of modeling its growth with an aim to design and control film's surface morphology. Oftentimes, smooth surface is desirable and is experimentally achievable via molecular-beam epitaxy (MBE) growth technique with exceptionally low deposition flux. Adatoms on the film grown with such a method tend to have large diffusion length which can be computationally very costly when certain statistical aspects are demanded. We present a graph theoretic approach to modeling MBE grown thin film with long atomic mean free path. Using Markovian assumption and given a local diffusion bias, we derive the transition probabilities for a random walker to traverse from one lattice site to the others after a large, possibly infinite, number of hopping steps. Only computation with linear-time complexity is required for the surface morphology calculation without other probabilistic measures. The formalism is applied to simulate thin film growth on a two-dimensional flat substrate and around a screw dislocation under the modified Wolf--Villain diffusion rule. A rectangular spiral ridge is observed in the latter case with a smooth front feature similar to that obtained from simulations using the well-known multiple registration technique. An algorithm to compute the inverse of a class of sub-stochastic matrices is derived as a corollary.
\end{abstract}


\maketitle

\section{Introduction}
Dated back to the classic Seven Bridges of K{\"o}nigsberg problem~\cite{BiggLloyWils76}, graph theory has been at the core of mathematics and computer science and has recently fueled considerable interests in understanding the behaviors of complex networks~\cite{Newman10, *Strogatz01, *AlbeBara02, *BoccLatoMoreChavHwan06}. One of the theory's vast applications is in the development of random walks on graphs~\cite{[{See for example: }] Lova93, *Chung97}. Two vertices, representing different system states, are linked by an edge if there is a non-zero transition probability connecting the two. This framework lends itself naturally to modeling particle diffusion on a substrate such as those observed in epitaxial growth of thin films~\cite{[{See }] [{ and references therein.}] OuraLifsSaraZotoKata03}. An adatom, after deposition, can hop from one adsorption site to the next by thermal excitation. The adatom will continue to hop as long as it can overcome the local potential barrier, after which it becomes a part of the substrate. Provided that the deposition rate is so low that one movable adatom is present at a given time, one can take its position as a system state. An edge connecting vertex $i$ to $j$ thus signifies a probability of the adatom to make a hop correspondingly from position $i$ to $j$. 
If there exists an equal probability for hopping from $j$ to $i$ for any pair of vertices $(i,j)$, then the graph is said to be undirected and the corresponding Markov chain is time-reversible. The probability of finding the particle at a later time is essentially governed by Laplace's equation in the continuum limit. In a more realistic system a particle may react to the anisotropy of its surrounding~\cite{Kellogg94, *Zhu98} which results in a graph with edges being weighed differently. For example, due to energy minimization, adatom tends to be attracted toward one of the existing islands on the substrate causing it to enlarge. This type of scheme is responsible for patterning on thin film surfaces, which prompts the question: \emph{what is the most probable surface pattern given the lattice structure and local diffusion rule?} 

The simplest realizations of modeling solid-on-solid layer-by-layer growth simulate the biased flow of atoms through a potential landscape by simple bond counting. In Wolf--Villain (WV) model~\cite{WolfVill90}, adatom moves in the direction of \emph{maximizing} its coordination number in the next discrete time step; while in Das Sarma--Tamborenea (DT) model~\cite{SarmTamb91}, it moves to \emph{increase} the number provided that the current one is not sufficiently high. Despite their simplicity, they are found to yield consistent simulation results with more realistic finite-temperature model using Arrhenius hopping rate~\cite{SarmLancKotlGhai96, *TambSarm93} and low-temperature MBE experiments~\cite{PalaKrim94}. Often these models are applied in a limited-mobility regime where adatom moves one lateral step in the direction prescribed by the model and comes to a halt. Naive extension of these models to mimic a large diffusion length by repeated applications of one-step motion cannot avoid prescribing an ad-hoc maximum cutoff distance, and often is too computationally intensive for problems which require large-scale and/or long-time simulations. In this regard, we choose to apply our formalism to simulating the spiral surface growth around a screw dislocation commonly observed in MBE grown films with lattice mismatch at the film-substrate interface such as that of GaN based devices~\cite{HeyiTarsElsaFini99, *ParkDabiBenjCohe01, *CuiLi02}, or certain semiconductor materials~\cite{SpriUetaFranBaue96}, and provides a mechanism for driving the growth of a class of nanowires~\cite{MoriJin10}.
Unlike the spiral growth in the limit of fast desorption, where the ridge motion can be determined locally and is well described by Burton--Cabrera--Frank (BCF) model~\cite{BurtCabrFran51, *CabrLevi56}, the minimal desorption limit where particle's diffusion length is comparable to the system size still presents a modeling challenge. Theoretical and computational investigations of spiral growth in this regime are typically carried out in a continuum limit using a phase-field method~\cite{RediRickKuhnRatz08, *YuLiuVoig09}. This method although provides an analytical handle to the problem, it suffers from the shortcomings of a continuum formulation, e.g., when accounting for system's anisotropy. Kinetic Monte-Carlo (kMC) method has also been chosen to explore the spiral growth~\cite{XiaoIwanAlexRose91, *XiaoIwanAlexRose91b}.  In order to suppress microscopic noise, either the use of multiple-registration~\cite{XiaoIwanAlexRose88, *Tang85} or atomic evaporation~\cite{XiaoIwanAlexRose91, *XiaoIwanAlexRose91b} schemes must be implemented. The latter, though more realistic, is less practical at the extremely low supersaturation limit being considered.

The paper is organized as follows. The mathematical structures will be discussed in Sec.~\ref{S:MathCon} where film's sites will be classified (Sec.~\ref{S:Classified}) and the limiting lattice transitions calculated (Sec.~\ref{S:LimitTrans}). Implementation and graph algorithms shall be mentioned in Sec.~\ref{S:GraphAlgo}. The formalism is applied using the extended Wolf--Villain model on a two-dimensional film growth on a flat substrate (Sec.~\ref{S:Flat}) and around a single screw dislocation (Sec.~\ref{S:Screw}). Results shall be discussed in Sec.~\ref{S:Results}, followed by concluding remarks in Sec.~\ref{S:Conclusion}

\section{Mathematical construct}\label{S:MathCon}
The mathematical question underlying our modeling of particle diffusion with possibly infinite diffusion length amounts to finding the probability that a random walker beginning its journey at position $i$ on a lattice would reach point $j$ by traveling along the edges connecting them. Edges linking, say, vertex $i$ to its neighbors are not necessary of equal weights. Biased or asymmetric walks are accounted for by having directed weighted edges. This type of walks is required for modeling diffusion in the presence of fields such as in electromigration~\cite{AntcEhrl08} or when adatoms sensitively react to local atomic configurations. The Markovian assumption that each step the walker takes is independent of the previous one was proven to be quite accurate~\cite{WangWrigEhrl89}. If only nearest-neighbor hopping is allowed, only edges linking nearest neighbors are present and the graph structure matches the structure of the lattice surface. In principle one can incorporate long atomic jumps by including edges linking remote sites with a suitably weight factor. This type of jumps has been observed experimentally with fairly significant jump rates~\cite{AntcEhrl07, *OhKohKyunEhrl02, *LindHorcLaegSten97}. For simplicity we shall only limit ourselves to one random walker. This allows us to easily define system's state at a given time by the lattice position of the walker. A more complicated scheme is required to define a state should one choose to model, e.g., clustered or collective diffusion~\cite{OuraLifsSaraZotoKata03}.

\begin{figure}[htb]
	\centering
	\includegraphics[width=.48\textwidth]{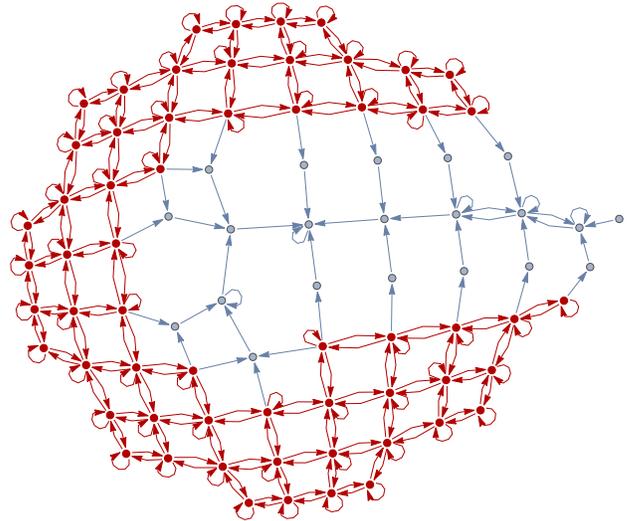}
	\caption{A graph representation of a transition matrix whose vertices symbolize states of the system in terms of adatom's position. Each arrow depicts an edge linking two positions whose transition probability from one to the other is non-zero. The probability is derived from WV diffusion rule on a circularly shaped substrate with a screw dislocation. One can see the trace of the underlying simple cubic lattice structure from the graph. Vertices and edges in red show a part of the graph which is strongly connected. These vertices form a class.}
	\label{fig:BeforeDAG}
\end{figure}
In this section, we shall outline the relevant portions of discrete time Markov chain needed for the modeling and state our graph algorithm for computing the infinite-hop transition matrix. For an illustrative purpose, we shall phrase the problem in the language of crystal growth where an adatom wanders from point $i$ to point $j$ according to a given set of diffusion rules by hopping between atomic positions (vertices) linking $i$ and $j$. Let a lattice configuration, some lattice-dependent property which influences the probability of the adatom such as atomic heights, be described by $\vec{H} \equiv \{H_i\}$. Define a one-step transition matrix $\Tmat(\vec{H})$ whose element $\Tran_{ik}(\vec{H})$ gives a transition probability for an atom at $i$ to go to one of the connecting sites $k$ which usually is a nearest neighboring lattice site. A non-zero element $\Tran_{ik}$ is thus equivalent to having a directed edge with a corresponding probabilistic weight connecting vertex $i$ to vertex $k$ of a graph as shown, for example, in Fig~\ref{fig:BeforeDAG}. We also account for in-place hopping through the matrix element $\Tran_{ii}$ (represented graphically by a vertex with self loop). Since the adatom must go somewhere $\sum_k \Tran_{ik} = 1$ which makes $\Tmat$ a stochastic matrix. Knowing $\Tmat$, one can easily find the transition probability from $i$ to $j$ if the atom takes \emph{exactly} $n$ hops using 
\begin{equation}\label{E:Pijnsteps}
	\Pn{n}_{ij} = \sum_{k} \Tran_{ik}\Pn{n-1}_{kj} = \!\!\sum_{k_1,k_2,\ldots, k_{n-1}} \!\! \Tran_{ik_1}\cdots \Tran_{k_{n-1}j}\,,
\end{equation}
or simply, $\Pmatn{n}=\Tmat^n$ in matrix notation. Since we are interested in the long diffusion limit, we shall investigate in particular the case where $n\to \infty$.

\subsection{Lattice site classification}\label{S:Classified}
Given the state of height configuration $\vec{H}$, one can classify lattice position $j$ (or system's state) according to the limiting behavior of $\Tmat^\infty$ into \emph{transient class} (denoted by $\Tset$) or \emph{recurrent class} (denoted by $\Rset$). In case of finite number of system's states~\cite{GrimStir92} (which is particularly relevant for modeling Markov processes on a computer), $j$ is transient if $\Prob^\infty_{ij} = 0$ and recurrent otherwise~\footnote{In particular these recurrent states are non-null. For $j \in \Rset$, it is non-null if $\mu_j \equiv \sum_{n=0}^\infty n F^n_{jj}$ is finite. Here $F^n_{ij}$ is the first passage probability which denotes the likelihood that a walker starting at $i$ will end up at $j$ for the first time after exactly $n$ hops. Thus $\mu_j$ gives mean recurrence time of state $j$ or the expected value of the time of the first visit to $j$ from $j$.}. Loosely speaking, if one lets an adatom wander for a very long time, it will end up at one of the recurrent sites. It is easy to see that there must be at least one such site, otherwise $\Prob^\infty_{ij} = 0$ for all $j$ with edge(s) directed from $i$. This would lead to $\sum_j \Prob^\infty_{ij} = 0$ and we would reach a contradiction because the adatom starting from $i$ must go somewhere. The real merit of this classification is that it can be done through permuting rows and columns of $\Tmat$ which amounts to relabeling of lattice positions, and prior to the actual computation of $\Tmat^\infty$ itself.

If $\Tmat$ is reducible, i.e., if there exists at least one transient site, then by definition one can find a non-unique permutation matrix $\Qmat_1$ that transforms $\Tmat$ into a block triangular form such that
$
	\Qmat_1^\top \cdot \Tmat\cdot \Qmat_1 = \left(\begin{smallmatrix} \mathds{X} & \mathds{Y} \\ \Zero &\mathds{Z} \end{smallmatrix}\right),
$
where $\mathds{X}$ and $\mathds{Z}$ are square matrices, and $\Zero$ is a matrix with all elements being zero. If $\mathds{X}$ and $\mathds{Z}$ are reducible still, then we can apply another symmetric permutation to them. If during the process there exist rows with nonzero entries only in diagonal blocks, these rows can be permuted to the bottom of the modified transition matrix. Finally we shall arrive at the upper-triangular block form or the \emph{canonical form for reducible matrices}~\cite{Meye00} of a system via permutation matrix $\Qmat$ formed from the products of previous permutation matrices:
\begin{equation}\label{E:Tperm}
	\Tmatperm = \Qmat^\top \cdot\Tmat\cdot\Qmat = \begin{pmatrix} \Tmattt & \Tmattr \\[0.3em]
											 \Zero & \Tmatrr \end{pmatrix},
\end{equation}
where
\begin{gather}\label{E:Tpermstructure}
	\Tmattt = \begin{pmatrix} \Transient_1 & \Transient_{1,2} & \cdots & \Transient_{1,t} \\
							  \Zero & \Transient_2 & \cdots & \Transient_{2,t} \\
							 \vdots & \ddots & \ddots & \vdots \\
							 \Zero & \cdots & \Zero & \Transient_t \end{pmatrix}, \\
	\Tmattr = \begin{pmatrix} \Trantorec_{1,1} & \cdots & \Trantorec_{1,r} \\
							 \vdots & \ddots & \vdots \\
							 \Trantorec_{t,1} & \cdots & \Trantorec_{t,r} \end{pmatrix}, 
	\Tmatrr = \begin{pmatrix} \Recurrent_1 &  & \Zero \\
							 & \ddots & \\
							 \Zero & & \Recurrent_r \end{pmatrix}. \notag			 
\end{gather}
Each diagonal block matrix $\Transient_1,\ldots,\Transient_t$ is either irreducible or $\Zero$ and $\Recurrent_1,\ldots,\Recurrent_r$ are irreducible and stochastic. In the language of graphs, the graph representation of matrix $\mathds{A}$ is irreducible if there is a sequence of directed edges linking every pair of vertices together, i.e., the graph is \emph{strongly connected}. As an example, the vertices and edges highlighted in red in Fig.~\ref{fig:BeforeDAG} make up a subgraph which is strongly connected. Thus their transition matrix representation is irreducible.

Through the application of $\Qmat$, the newly ordered, one-step transition matrix $\Tmatperm$ shows the separations of lattice sites into $t$ transient classes $\Tset_1, \ldots, \Tset_t$, and $r$ recurrent classes $\Rset_1, \ldots, \Rset_r$. The form also effectively suggests from which transient classes is a site in a recurrent class \emph{accessible}~\footnote{Mathematically speaking, site $j$ is said to be accessible from $i$ ($i\access j$) if there is a non-zero probability that, starting from $i$, the adatom will reach $j$ at some future hop ($\Pn{n}_{ij} > 0$ for some $n \ge 0$). Sites $i$ and $j$ communicate with each other ($i \leftrightarrow j$) if and only if $i\access j$ and $j\access i$, which can happen only when both $i$ and $j$ belong to the same class.}. It is therefore of paramount importance to devise an algorithm to construct $\Qmat$. We shall postpone this discussion to Section~\ref{S:GraphAlgo}.

\subsection{Limiting lattice transitions}\label{S:LimitTrans}
To arrive at the limiting probability transition matrix $\Tmat^\infty$, one is interested in examining the possibility of transitions between elements in $\Tset$ and/or $\Rset$. We shall state without proofs these results~\cite{[{For rigorous proofs, please consult, e.g., }] Meye00}, some of which are most evident from the structure of $\Tmatperm$ in Equations (\ref{E:Tperm}) and (\ref{E:Tpermstructure}). The following transitions from $i$ to $j$ lead to vanishing probability, $\Prob^\infty_{ij} = 0$: (a) $i, j \in \Tset$, (b) $i\in \Rset_1$ but $j\in \Rset_2$, and (c) $i \in \Rset$ and $j \in \Tset$. In other words, after a large number of hops, if a random walker beginning its trip from any of the transient classes, it will eventually go to a recurrent class. If, however, it already starts its journey in one of the recurrent classes, it will stay there forever. The remarkable theorem due to Oskar Perron and Georg Froebenius~\cite{Perr07, *Frob12} enables us to compute the limiting probability distribution of the latter case when it exists, and otherwise provides the fraction of time the walker spends on each site in the class.

When a random walker starts its trip within a recurrent class, say $\Rset_f$ with $m$ members, whose transition matrix is given by $\Recurrent_f$, that the infinite-hop limit of the transition matrix $\Recurrent_f^\infty \equiv \lim_{n\to\infty} \Recurrent_f^n$ exists depends on the periodicity of $\Recurrent_f$. Since $\Recurrent_f$ is irreducible, it is \emph{aperiodic} if there is only one eigenvalue with modulus 1 and the limit exists, otherwise it is \emph{periodic} and the limit does not exist. A good example of a periodic matrix is $\bigl( \begin{smallmatrix} 0 & 1 \\ 1 & 0 \end{smallmatrix} \bigr)$, where each self multiplication gives the result which fluctuates between itself and the identity matrix. Physically speaking, if a walker enters this class, its position at later times will oscillate between one of the two sites indefinitely. The period of this chain is therefore two. More generally the period coincides with the number of eigenvalues of modulus 1~\footnote{Alternatively one can find the period $p$ from the characteristic equation of $\Recurrent_f$ without directly computing its eigenvalues. See Ref.~\onlinecite{Meye00} for proof.}.
Once in a class, the probability distribution of the walker's positions is not constant in time but continually progresses from subclass to subclass and eventually returns to the original distribution after going through all $p$ subclasses.
In practice classifying a recurrent class by its periodicity is simple; a non-negative irreducible matrix is aperiodic if there is at least one positive element along the diagonal~\cite{Meye00}.

When the limit exists,
\begin{equation}\label{E:PerronProj}
	\Recurrent_f^\infty = \evec_{f}\pivec_{f}^\top = 
	\begin{pmatrix} 
		\pi_{f_1} & \pi_{f_2} & \cdots & \pi_{f_m} \\
		\pi_{f_1} & \pi_{f_2} & \cdots & \pi_{f_m} \\
		\vdots & \vdots & & \vdots \\
		\pi_{f_1} & \pi_{f_2} & \cdots & \pi_{f_m}
	\end{pmatrix},
\end{equation}
where $\evec_{f}$ is a vector whose elements are all 1, and $\pivec_{f}$ is a unique Perron vector satisfying $\Recurrent_f^\top\cdot \pivec_{f} = \pivec_{f}$, with all positive elements and properly normalized so that $\|\pivec_{f}\|_1 = 1$. This vector represents the distribution of probabilities that the walker will be at a particular site eventually. On the other hand, when $\Recurrent_f$ is periodic, $\evec_{f}\pivec_{f}^\top$ in (\ref{E:PerronProj}) is the solution to the Ces\`aro average, i.e.,
\begin{equation}\label{E:CesaroAvg}
	\Cesaro_f \equiv \lim_{n\to\infty} \frac{\Id + \Recurrent_f + \Recurrent_f^2 + \ldots + \Recurrent_f^n}{n} = \evec_{f}\pivec_{f}^\top.
\end{equation}
Matrix element $[\Cesaro_f]_{ij}$ represents the portion of time that the walker hops onto $j$ irrespectively of its starting position $i$. 
Henceforth, for the sake of theoretical discussion, whenever we examine infinite-hop probability within a recurrent class, we shall adopt the Ces\`aro average interpretation in place of $\Recurrent_f^\infty$ when the latter does not exist. 

The remaining question is to determine how probable it is for a walker to end up in one of the above recurrent classes if it starts from a transient class. Suppose the hop starts from a site in $\Tset_s$, the walker might have to visit subsequent intermediate transient classes $\Tset_m$'s, for all $m$ connecting $s$ to recurrent class $\Rset_f$. The total probability will ultimately involve how long the walker spends in each class as it traverses. Let matrix element $[\Mmat_s]_{ij}$ denotes the expected total number of hops onto site $j\in \Tset_s$ given that the first hop starts at $i\in \Tset_s$, and $\Transient_s$ be the transition probability matrix among members in $\Tset_s$. Since the walker will either hop onto $j$ with probability $[\Transient_s^n]_{ij}$ in which case the hop value is 1, or it won't which brings the hop value to 0. This means $[\Transient_s^n]_{ij}$ also represents the expected number of hop the walker will step onto $j$ on the $n^\text{th}$ step.
Thus the total number of times the walker steps onto $j$ on average is calculated from the total contributions from all steps:
\begin{equation}\label{E:Mmat}
	\Mmat_s = \sum_{n=0}^\infty \Transient_s = (\Id - \Transient_s)^{-1}
\end{equation}
The non-negativity and irreducibility of $\Transient_s$, and the fact that all of its eigenvalues have modulus strictly less than 1 ensure that the above Neumann series exists and is positive definite~\cite{Meye00}. Matrix $\Id - \Transient_s$ is an example of what's called \emph{M-matrix}, and often emerges in relation to systems involving linear or nonlinear equations in many areas including solving finite difference methods, problems in operations research, and Markov processes~\cite{BermPlem94}.

Consider a transient class $\Tset_{s+1}$ which can be reached only from $\Tset_s$. The probability that a walker starting at site $i \in \Tset_s$ will wander to $j \in\Tset_{s+1}$ after an infinite number of hops must equal the expected period that the walker is going to spend on some site $k \in \Tset_s$ times the probability that it will exit $\Tset_s$ through $k$ into $\Tset_{s+1}$, summing over all transitory sites $k$'s:
\begin{equation}\label{E:Pexit}
	\sum_{k\in \Tset_s} [\Mmat_s]_{ik}[\Transient_{s,s+1}]_{kj} = [\Mmat_s\cdot\Transient_{s,s+1}]_{ij}
\end{equation}
It is easy to extend the result in (\ref{E:Pexit}) to the case where there are other intermediate transient classes and/or more than one route for the walker to take until it reaches some chosen transient class $\Tset_x$. Let $\pvec = \{p_1,\ldots,p_m\}$ be a path that connects transient class $\Tset_s$ to $\Tset_x$ via $m$ intermediate classes $\Tset_{p_1}, \ldots, \Tset_{p_m}$. The exiting probability matrix is given by summing over the contributions from all such paths according to
\begin{equation}\label{E:Pexittotal}
	\Pexit{s}{x} = \sum_\pvec \Mmat_s\cdot\Transient_{s,p_1}\cdot\Mmat_{p_1}\cdot\Transient_{p_1,p_2}\cdots \Mmat_{p_m}\cdot\Transient_{p_m,x}.
\end{equation}
Moreover the probability that the walker in $\Tset_x$ will transit to $\Rset_f$ is simply $\Mmat_x\cdot\Trantorec_{x,f}$. Combining this result with (\ref{E:Pexittotal}), we finally arrive at the expression for the probability that a walker starting from a site in $\Tset_s$ will be entrapped in $\Rset_f$ after a large number of hops:
\begin{equation}\label{E:TranToRecProb}
	\Trantorec^\infty_{s,f} = \sum_x \Pexit{s}{x}\cdot\Mmat_x\cdot\Trantorec_{x,f}\cdot\Recurrent_f^\infty
\end{equation}
The sum $\sum_x$ is taken over all possible $\Tset_x$'s from which $\Rset_f$ can be accessible. 

The result of this analysis can be summarized by the following matrix,
\begin{equation}\label{E:Pinfinity}
	\Tmatperm^\infty = \begin{pmatrix}
		\Zero & \cdots & \Zero &\Trantorec^\infty_{1,1} & \cdots & \Trantorec^\infty_{1,r} \\
		\vdots &	 \ddots	& \vdots & \vdots & \ddots & \vdots \\
		\Zero &	\cdots	& \Zero & \Trantorec^\infty_{t,1} & \cdots & \Trantorec^\infty_{t,r} \\
		\Zero & \cdots & \Zero & \Recurrent^\infty_1 & & \Zero \\
		\vdots &	 \ddots	& \vdots & 	 & \ddots & \\
		\Zero & 	\cdots	& \Zero & \Zero & & \Recurrent^\infty_r
	\end{pmatrix},
\end{equation}
together with $\Trantorec^\infty_{s,f}$ as defined in (\ref{E:TranToRecProb}), and $\Recurrent_f^\infty$ as in (\ref{E:PerronProj}) with the appropriate limit interpretation. Given initial state vector $\vvec_i$, after a large number of hops, the system's probability distribution is therefore
\begin{equation}
	\vvec_f = \vvec_i\cdot\Tmatperm^\infty.
\end{equation}

\subsection{Graph algorithms}\label{S:GraphAlgo}
The analysis so far has made full use of the canonical form for reducible matrices. The question thus arises: is there a way to systematically find a permutation which would cast a reducible matrix into its canonical form? Fortunately in graph theory there exists a set of algorithms which does exactly this. Initially one can find a permutation which could swap indices in such a way that strongly connected components (or lattice sites in this case) are grouped together into appropriate classes. 
Algorithms such as Tarjan's and Gabow's exist to do this with linear-time complexity~\cite{AspvPlasTarj79, *CormLeisRiveStei01}. At this point, the relationship between classes can be represented by a non-unique \emph{Directed Acyclic Graph} (DAG). 

\begin{figure}[htb]
	\centering
	\includegraphics[width=.5\textwidth]{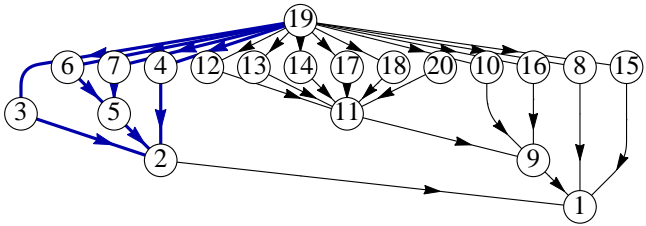}
	\caption{The directed acyclic graph (DAG) of the graph shown in Fig.~\ref{fig:BeforeDAG}. Each vertex represents a class of strongly connected components, and each edge connects two classes with non-vanishing transition probability from one to the other. Vertex $19$, for example, represents the subgraph in Fig.~\ref{fig:BeforeDAG} that is highlighted in red.}
	\label{fig:DAG}
\end{figure}
Let $G= (V,E)$ be a DAG with a set of vertices $V=\{1,2,\ldots,t+r\}$ and a set of edges $E$. Vertex $i$ represents a transition matrix of type (a) $\Transient_i$ for all $i$ whose outdegree is positive ($\deg^+(i) > 0$); or (b) $\Recurrent_i$ if $i$ is a \emph{sink} ($\deg^+(i) = 0$). The underlying graphs of these matrices are, by construction, strongly connected which make them irreducible. Edge $(i,j)$ connecting vertex $i$ to $j$ represents transition matrix of type (a) $\Transient_{i,j}$ from $\Tset_i$ to $\Tset_j$ if $\deg^+(j) > 0$; or (b) $\Trantorec_{i,j}$ from $\Tset_i$ to $\Rset_j$ if $j$ is a sink. 
An example of such a graph is shown in Fig.~\ref{fig:DAG}.
Vertices in a DAG have a natural ordering which could best be visualized by a layered tree, pointing one way from vertices of transient classes to those of recurrent classes. A \emph{topological sorting}~\cite{Kahn62, *Tarj76} can be performed in linear time ($\mathcal{O}(|V|+|E|)$) to yield a permutation of classes from transient to recurrent. By composing the two permutations together, one obtains a permutation which takes a reducible matrix to its canonical form.
In practice it is not important to topologically sort DAG to obtain the full canonical form to efficiently compute elements of $\Tmatperm^\infty$. The ordering of $\Mmat_a\cdot\Transient_{a,b}$ terms in Equation~(\ref{E:Pexit}) only requires that they appear in the same sequence as the underlying DAG. Topological sorting simply relabels the class numbers in order of appearance which yields no new information, and thus is unnecessary.

In cases where a prescribed diffusion rule prohibits self hopping ($P_{ii}\ne 0, \forall i$), the limiting probability transition may not exist for some recurrent class $\Rset_i$. If preferred, one can determine the period of $\Rset_i$ directly from its underlying subgraph $G^i$. From graph theoretic perspective, $G^i$ is periodic with period $p$ if and only if it can be partitioned into $p$ smaller graphs $G^i_1, \ldots, G^i_p$ such that (a) if vertex $m$ is in $G^i_k$ and an edge $(m,n)$ connects $m$ to $n$ then it's implied that $n$ is in $G^i_{(k+1)\hspace{-.5em}\mod\! p}$; and (b) $p$ is the largest possible integer with this property. This makes sure that each transition takes the walker to a new class before it returns to the original class after $p$ successive transitions. An aperiodic recurrent class is one where such partition is not possible. The proof of the above theorem and the graph algorithm for finding the period of these ``cyclically moving classes'' are given in Ref.~\onlinecite{JarvShie99}.

To employ $G$ in the limiting probability calculation, we start by assigning appropriate matrices to all vertices and edges. Let $S = \{j|\deg^+(j)=0 \}$ be the set of all sinks of $G$. We then assign matrix $\Mmat_i$ to each vertex $i \notin S$, or $\Recurrent_i^\infty$ for $i \in S$. Each edge $(i,j)$ is prescribed by transition matrix $\Transient_{i,j}$ for $j \notin S$, and $\Trantorec_{i,j}$ for $j \in S$. To compute the limiting probability that a walker would reach a site in one of the recurrent classes in $S$, we begin by giving the walker his initial probability distribution vector $\vvec_0$ at the starting class. This vector contains only one non-zero component of value 1 at the position corresponding to the dropped site. Then we scan the graph from the starting vertex (which may or may not be the source) to the sinks. As we traverse the graph to vertex $s$, we would have accrued all the probability contributions along that path prior to reaching $s$. The probability distribution $\vvec_s$ stored at each vertex as it's visited would be
\begin{equation}
	\vvec_s = \begin{cases} \sum_{r} \vvec_r\cdot\Mmat_r\cdot\Transient_{r,s} & \text{if}~s \notin S, \\
						 \big(\sum_{r} \vvec_r\cdot\Mmat_r\cdot\Trantorec_{r,s}\big)\cdot\Recurrent_s^\infty & \text{if}~s \in S.
			\end{cases}
\end{equation}
The above summation $\sum_{r}$ is taken over all incoming vertices $r$ that point toward $s$. In a special case where the graph is made up of just one vertex, the final distribution is simply the Perron vector $\pivec_s$. It should be emphasized that direct computations of all the $\Mmat_s$ are not necessary. One only needs to calculate $\xvec \equiv \vvec_r\cdot\Mmat_r$ which is equivalent to solving the system of linear equations of the form $(\Id-\Transient_r^\top)\cdot \xvec = \vvec_r$. There exists many iterative schemes to determine $\xvec$ such as Jacobi method and successive over-relaxation (SOR) method~\cite{BermPlem94, Varg00}, or one could solve them using the equivalent constrained minimization method.
As an illustration, according to the highlighted subgraph of $G$ as shown in Fig.~\ref{fig:DAG}, vertex $2$ would receive probability vector $\vvec_2$ whose value equals
\begin{align*}
	\vvec_2  &= 
	\vvec_{19}\cdot\Mmat_{19}\cdot\bigg[ \Transient_{19,3}\cdot \Mmat_{3}\cdot \Transient_{3,2} + \Transient_{19,4}\cdot \Mmat_{4}\cdot \Transient_{4,2} \\
	& + \big(\Transient_{19,6}\cdot \Mmat_{6}\cdot \Transient_{6,5}
	+ \Transient_{19,7}\cdot \Mmat_{7}\cdot \Transient_{7,5}\big)\cdot\Mmat_{5}\cdot\Transient_{5,2} \bigg].
\end{align*}
Vector $\vvec_2$ can be interpreted as the probability that a walker would reach each site in transient class $\Tset_2$ given its initial probability distribution of $\vvec_{19}$ in class $\Tset_{19}$. By the time sink $1$ is reached, the limiting probability distribution vector $\vvec_1$ can readily be obtained from the $\vvec_r$'s of its immediate predecessors:
\begin{multline*}
	\vvec_1 = \bigg[ \vvec_2\cdot\Mmat_2\cdot\Trantorec_{2,1} + \vvec_8\cdot\Mmat_8\cdot\Trantorec_{8,1} \\
	+ \vvec_9\cdot\Mmat_9\cdot\Trantorec_{9,1} + \vvec_{15}\cdot\Mmat_{15}\cdot\Trantorec_{15,1} \bigg]\cdot \Recurrent_1^\infty
\end{multline*}
Other aspects of graph algorithm will be discussed in the respective sections as we examine the modeling problems. Reader interested in seeing the connection between DAG and the matrix inversion of the form $(\Id - \Amat)^{-1}$ where $\Amat$ is sub-stochastic should take a look in Appendix~\ref{S:Inverse}.

\section{Modeling}\label{S:Model}
For simplicity we shall only look at simple cubic lattice with only one type of atom so that there is no lattice mismatch between the substrate and the film. 
Here we consider two examples: (1) film grown on an initially flat substrate, and (2) film grown around a screw dislocation. 
For each initial profile of the substrate, graph $g$, similar to what's shown in Fig.~\ref{fig:BeforeDAG}, is constructed where a vertex represents a lattice position and an edge links a pair of adjacent neighbors together. Each vertex contains a number representing the height of the stack of atoms at that site. This implies that overhangs and voids are prohibited. An edge $(i,j)$ contains probability $P_{ij}$ that if an atom is deposited at $i$, it would move to $j$. For illustrative purposes, we choose Wolf--Villain diffusion model to prescribe such weight. According to the model~\cite{WolfVill90}, an adatom will try to move in such a way that the lateral coordination number is maximized. Should there be more than one such directions, the probability is divided equally among them. We made a slight alteration to the rule by including the coordination number at the present position into consideration so that in-place hopping is possible ($P_{ii}\ne 0$). This modification permits the limiting transition probability $\Recurrent_s^\infty$ to always exist in our analyses. The underlying graph $g$ of the transition matrix $\Pmat$ serves as the starting point of all of our simulations.

Before we discuss problem-specific modelings, let us examine the structure of the sinks obtained from WV model. Most of the time, a sink class only contains one member. This member is the representation of a kink site. Thus the task of computing Perron vectors to represent the $\Recurrent_s^\infty$ matrices is removed. Essentially these matrices are simply $\{1\}$. In a few rare cases we could have a situation where two, three or four kinks facing each other creating an area of two to four sites whose coordination numbers equal one another, and are higher than those of their surrounding neighbors'. Each vertex within a class with two (four) elements has two (three) edges, one pointing to itself and the rest point(s) to its neighbor(s). The limiting recurrent matrix $\Recurrent_s^\infty$ in this case would be a $2\times 2$ ($4\times 4$) matrix with all elements being $1/2$ ($1/4$). In the case of three elements, there is one vertex with three outgoing edges, while the other two only contain two edges. The Perron vector used in Eq.~(\ref{E:PerronProj}) comprises two of $2/7$ and one of $3/7$. Only in this last case is the weight not distributed evenly and the walker would more likely go there. Fortunately these cases never crop up in the analyses of the problems considered here.

\subsection{Growth on flat substrate}\label{S:Flat}
Here we consider the growth of thin film on an initially flat rectangular surface with the periodic boundary conditions, and apply the algorithm discussed in the previous section to find the most likely site that each atom would like to be. Two different methods shall be used: (I) each deposition site is chosen randomly and the algorithm gives its most probable final position; and (II) the most probable final position of all initial sites shall be chosen. Notice that the first atom to be dropped onto the surface will as likely be at any one site as another. Thus, for a visual purpose, we put it at the center of the surface. This atom will act as the seed to which subsequent atoms can attach themselves in the process of island formation.

In Method I, for each iteration, the simulation scheme starts by randomly selecting a starting position $i_0$. Then only the subgraph of $g$ whose components can be reached from $i_0$ is extracted. This process helps keep only the relevant irreducible classes for future computations. Subsequently the DAG $G$ of this subgraph is constructed by grouping strongly connected components together. We choose the eventual resting position by looking for $j$ which yields the greatest $P_{i_0 j}$. From the structure of $G$, one or three things could happen: (a) there is only one class thus $G$ is the sink; (b) $G$ only contains one sink and $j$ will inevitably be in that sink; or (c) $G$ contains multiple sinks and further calculation needs to determine the most likely sink that $j$ would eventually lie. It is only this last case that an actual calculation in the form outlined in Sec.~\ref{S:GraphAlgo} is performed if all one wishes is to get the final position of the adatom without calculating any statistics. Once all $\vvec_s, \forall s\in S$ are obtained, the chosen sink is the one whose member has the greatest value among all elements in all sink classes. Should there be more than one such members, the chosen site is selected randomly from that list. We then increment the atomic height at that site by one unit and the whole routine is repeated. Method II is similar to Method I except for one important point; our initial probability distribution $\vvec_i$ is given by $(1/N)\{1,\ldots, 1\}$, instead of having one non-zero element at a random position $i_0$.

\subsection{Growth around a screw dislocation}\label{S:Screw}
In modeling the growth of thin film around a screw dislocation, we decide upon a circularly shaped substrate of radius $r$ with free boundary. This choice ensures that any rectangular pattern that might emerge from the growth is due to the underlying lattice structure and not due to the shape of the boundary. Sites along the rim of the disk only connects with those within; thus they contain less nearest neighbors than the ones within the disk. The total number of lattice positions is approximately $\pi r^2$. We first initialize the height of all lattice points according to 
$h(x,y) = (b/2\pi)\tan^{-1}(y/x)$. Traversing around the dislocation core at $(0,0)$ once in the clockwise (counter-clockwise) direction will result in a height increment (decrement) approaching $b$ at large distance from the core.
In order to specify the coordination number at each point on the lattice, one needs to specify the criterion for height difference between any adjacent sites. If the height difference between two nearest-neighboring sites is smaller than $0.2 b$, we consider them as living on the same plane, and an adatom on top of the shorter site will not receive the coordination number count from the taller one. 
The simulation procedures in this case are the same as that of the flat surface. After the most probable site is found in either Method I or Method II, we increase its height by $b$ to match the magnitude of the Burgers vector. Then the process is repeated.

\section{Results and discussions}\label{S:Results}

\begin{figure}[htb]
	\centering
    \includegraphics[width=.23\textwidth]{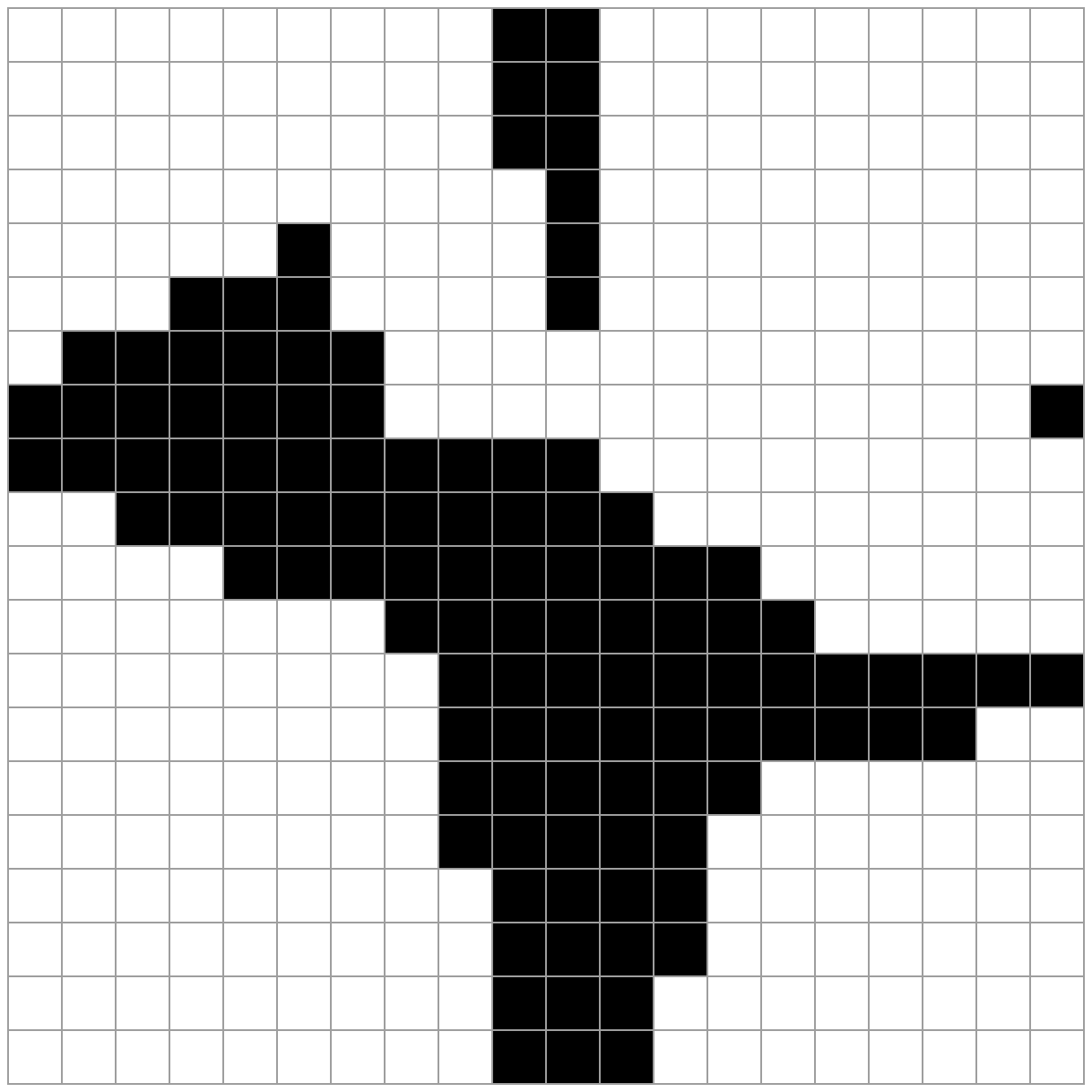}
    \includegraphics[width=.23\textwidth]{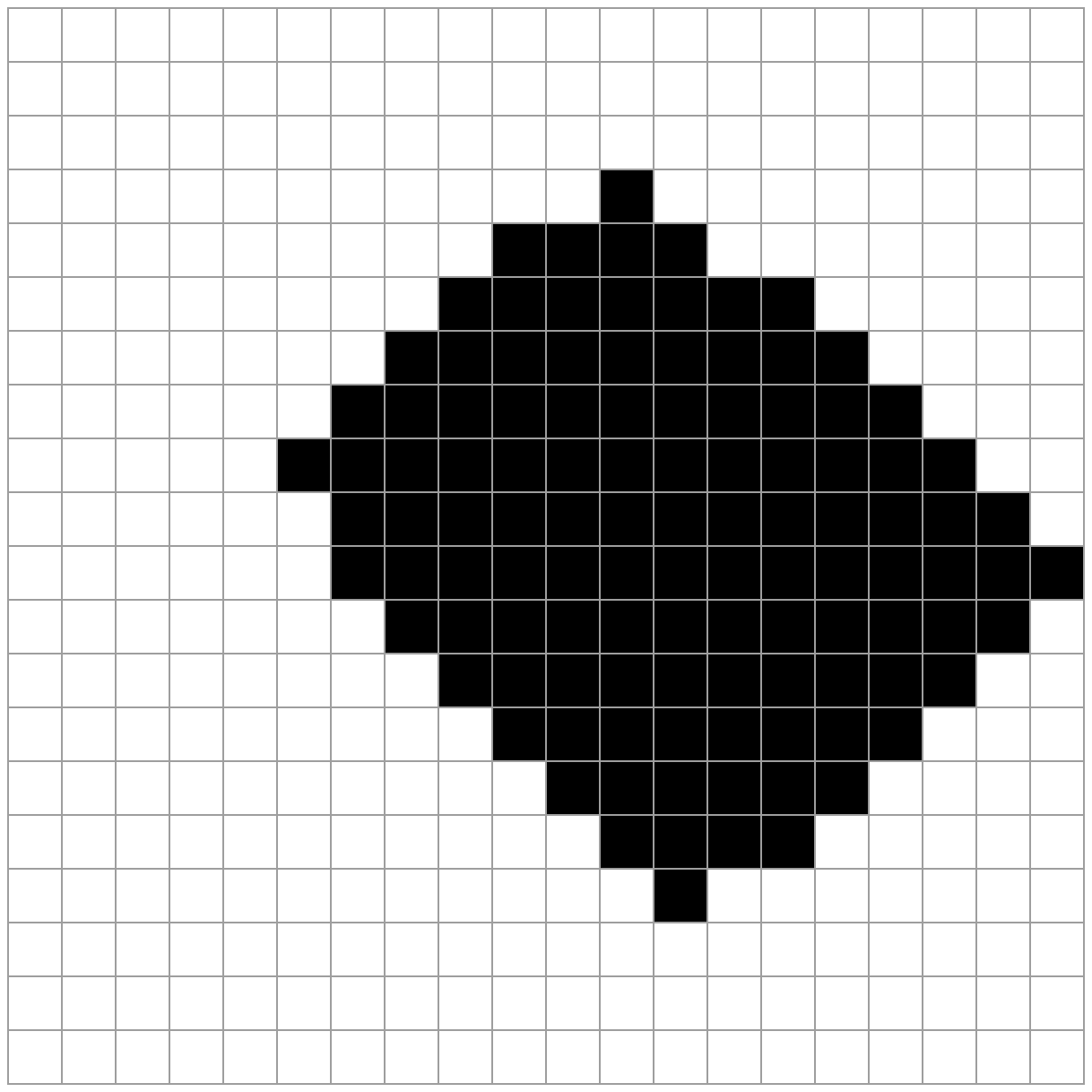}\\
    (a) \hspace{1.5in} (b)
    \caption{The film's surface on a periodic substrate of size $20\times 10$ lattice sites after 113 atoms are deposited for (a) Method I, and (b) Method II.}
    \label{fig:FlatSubstrates}
\end{figure}

The film's evolution in the case of growth on a flat substrate gives rise to one-island formation. In Method I, subsequent atoms attach themselves to the initial single-atom island since this is the only place where they can maximize their coordination number. The most probable final position of each newly deposited atom given by our algorithm tends to be the kink site that is closest to the dropped position. This results in an island surface with jagged island boundary as shown in Fig.~\ref{fig:FlatSubstrates} (a). The island will continue to grow until the first layer is completely filled up. This happens because there is no mechanism, e.g., Ehrlich--Schwoebel-like barrier, to prevent an adatom dropped on top of the island from hopping down to a lower layer where it could increase its coordination number. After the complete first layer is filled, the growth process repeats itself again and again, giving us a perfect layer-by-layer growth.

In Method II, the growth of the film is largely symmetrical. Identifying the plane of the substrate to be (001), atoms collectively tend to form an island whose boundaries grow outward in a rectangular fashion with growth fronts perpendicular to the [110], [1$\bar{1}$0], [$\bar{1}$10] and [$\bar{1}\bar{1}$0] directions as can be seen in Fig.~\ref{fig:FlatSubstrates} (b). Unlike the previous case, the shape of the island is generally very compact. The development of the growth fronts can be observed since the early stage in the simulation. These orientations are favorable to incoming adatoms than others since they provide more lateral kink sites which result in higher coordination numbers than if the front were one of [100], [010], [$\bar{1}$00] or [0$\bar{1}$0]. As in the previous method, the film grows one layer at a time.

The simulation result of the film grown around a screw dislocation shows a more interesting dynamics. We initially align the ridge so that it extends radially outward from the dislocation core at $(0,0)$ along the $[0\bar{1}0]$ direction. When viewed from the top, the ridge starts spiraling outward in the clockwise direction (since the atomic height difference is $h(x=0^+)-h(x=0^-) = +b$ along the ridge) as more and more atoms are attracted toward its left side. 
The film's evolution according to Method I, like the flat substrate case, gives a very rough spiral ridge. The randomness of each deposition causes the atomic incorporation to occur at the nearest kink site which may be anywhere. This makes it difficult to describe how the film grows generally. More importantly, the growth does not reflect the shape observed in actual experiments where the spiral ridge fronts are of well-defined compact geometrical form~\cite{HeyiTarsElsaFini99, *ParkDabiBenjCohe01, *CuiLi02}.

\begin{figure}[htb]
	\centering
    \includegraphics[width=.23\textwidth]{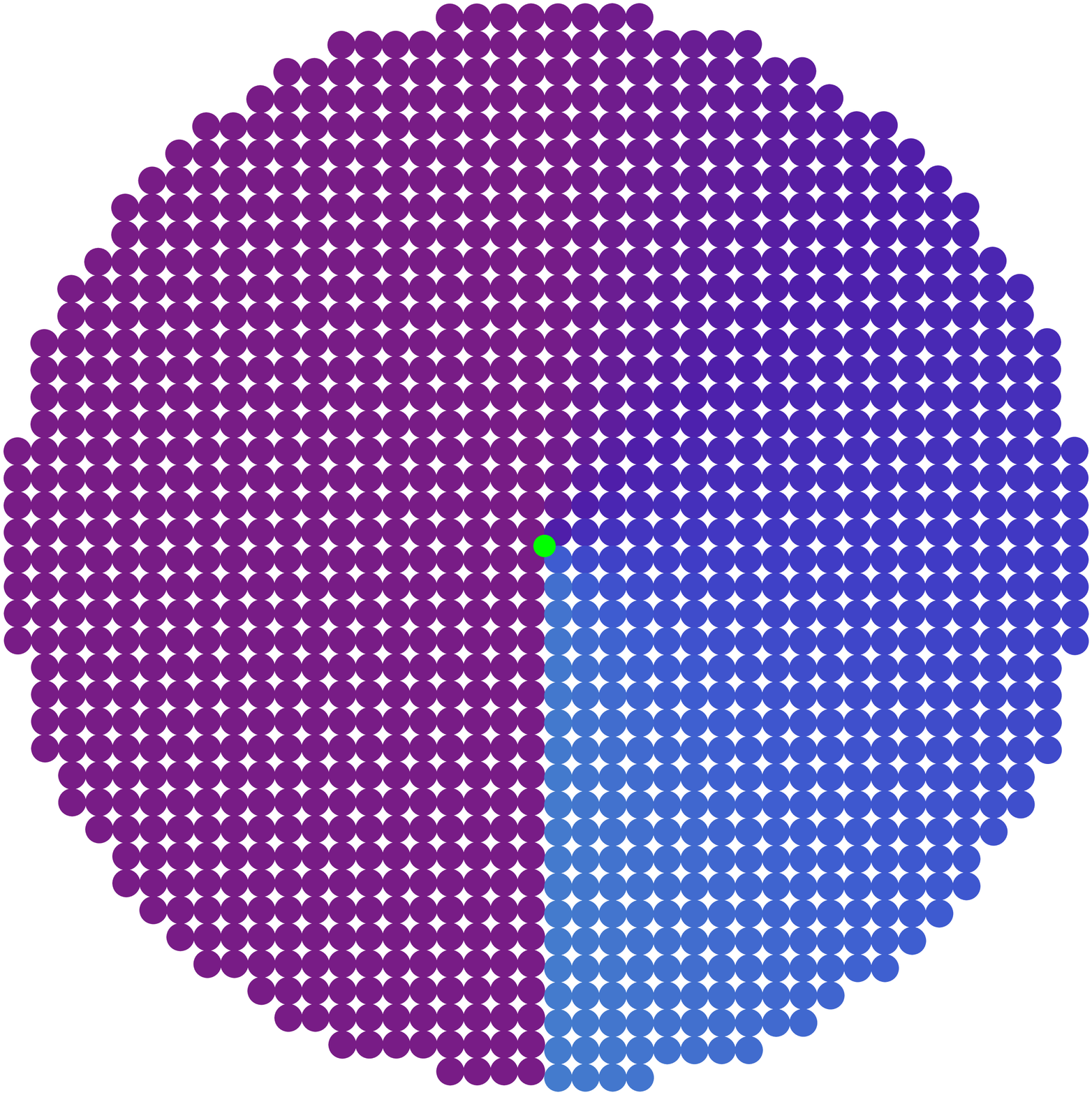}
    \includegraphics[width=.23\textwidth]{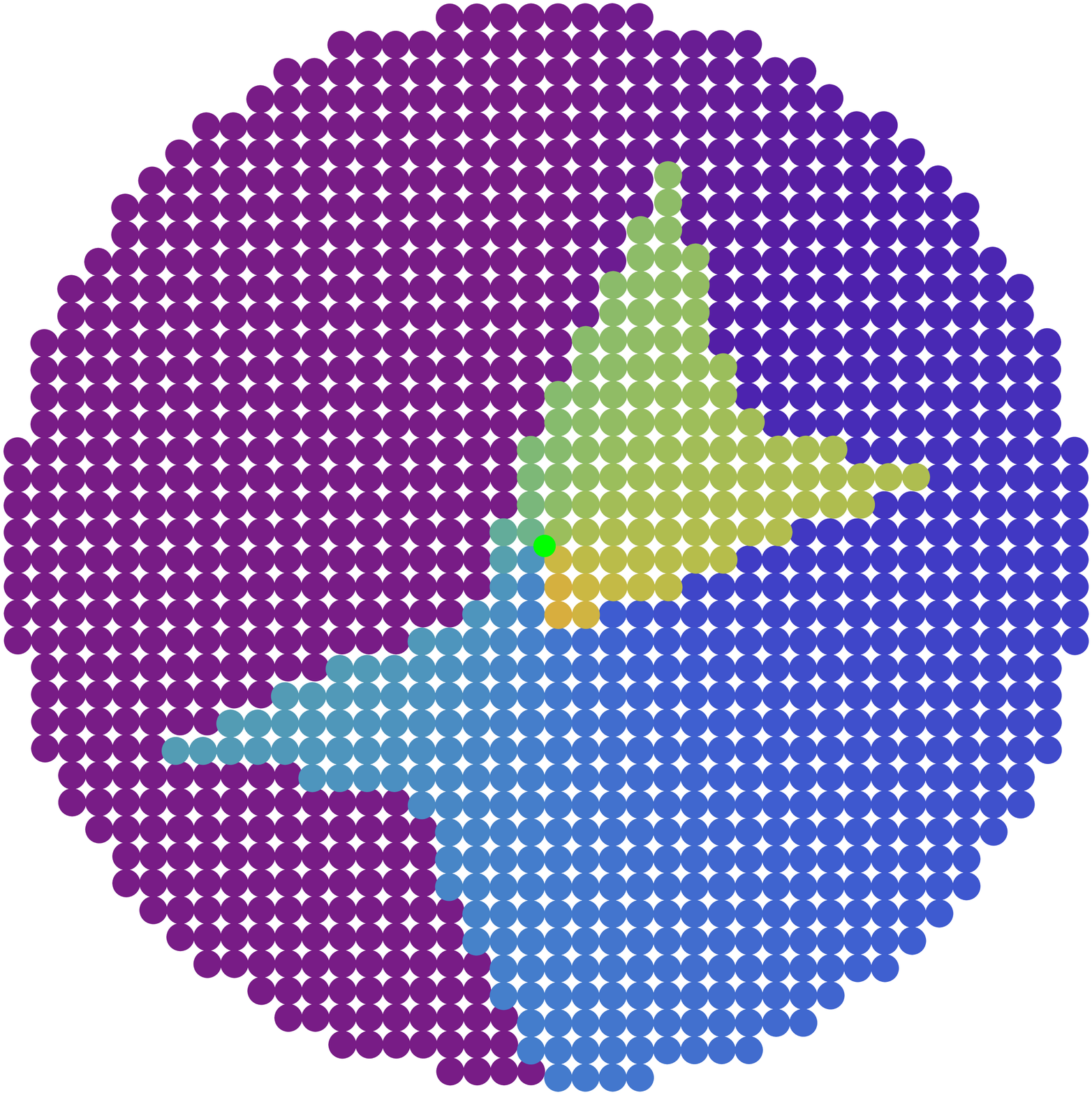}\\
    (a) \hspace{1.5in} (b)\\
    \vspace{.2in}
    \includegraphics[width=.23\textwidth]{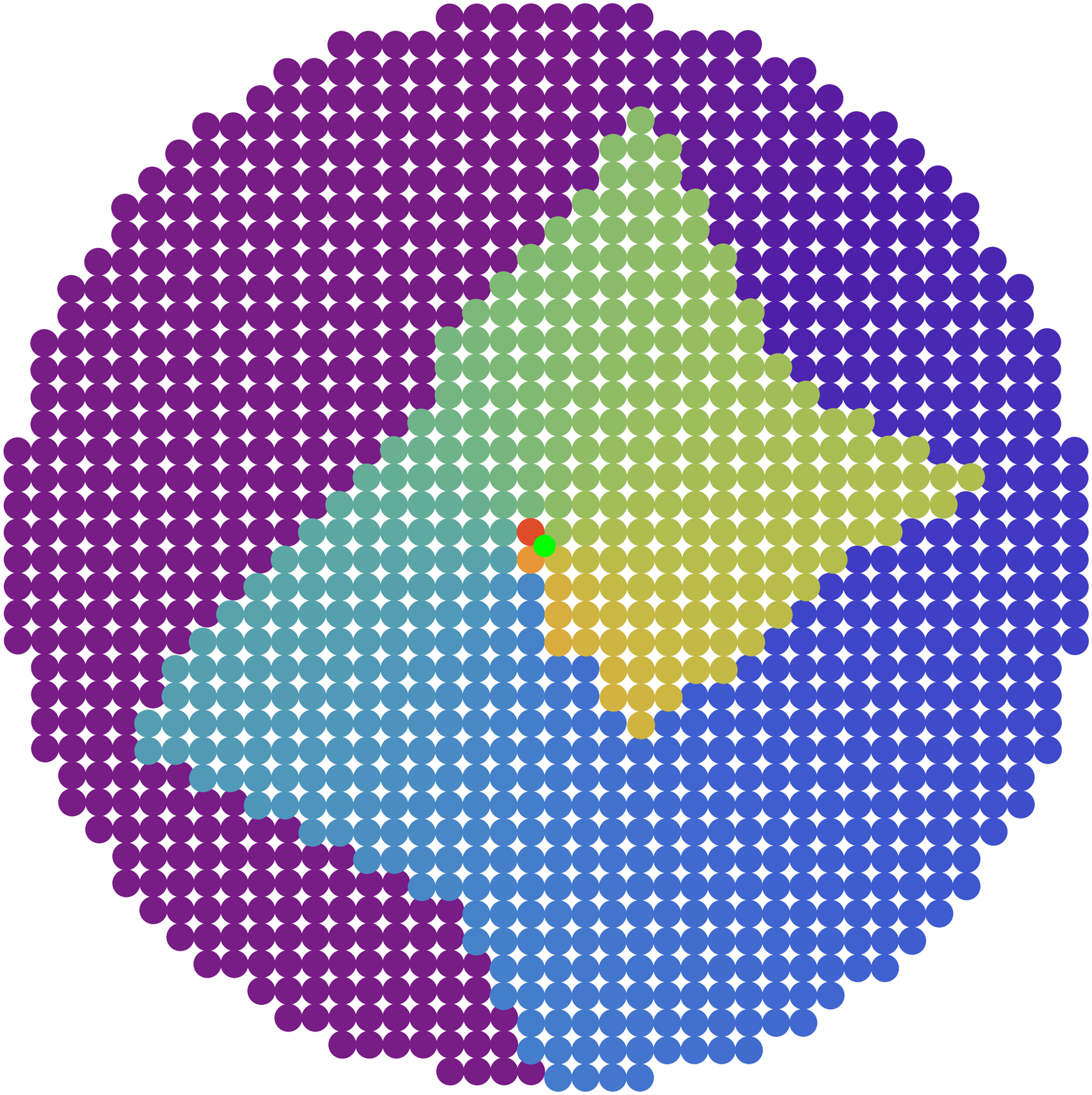}
    \includegraphics[width=.23\textwidth]{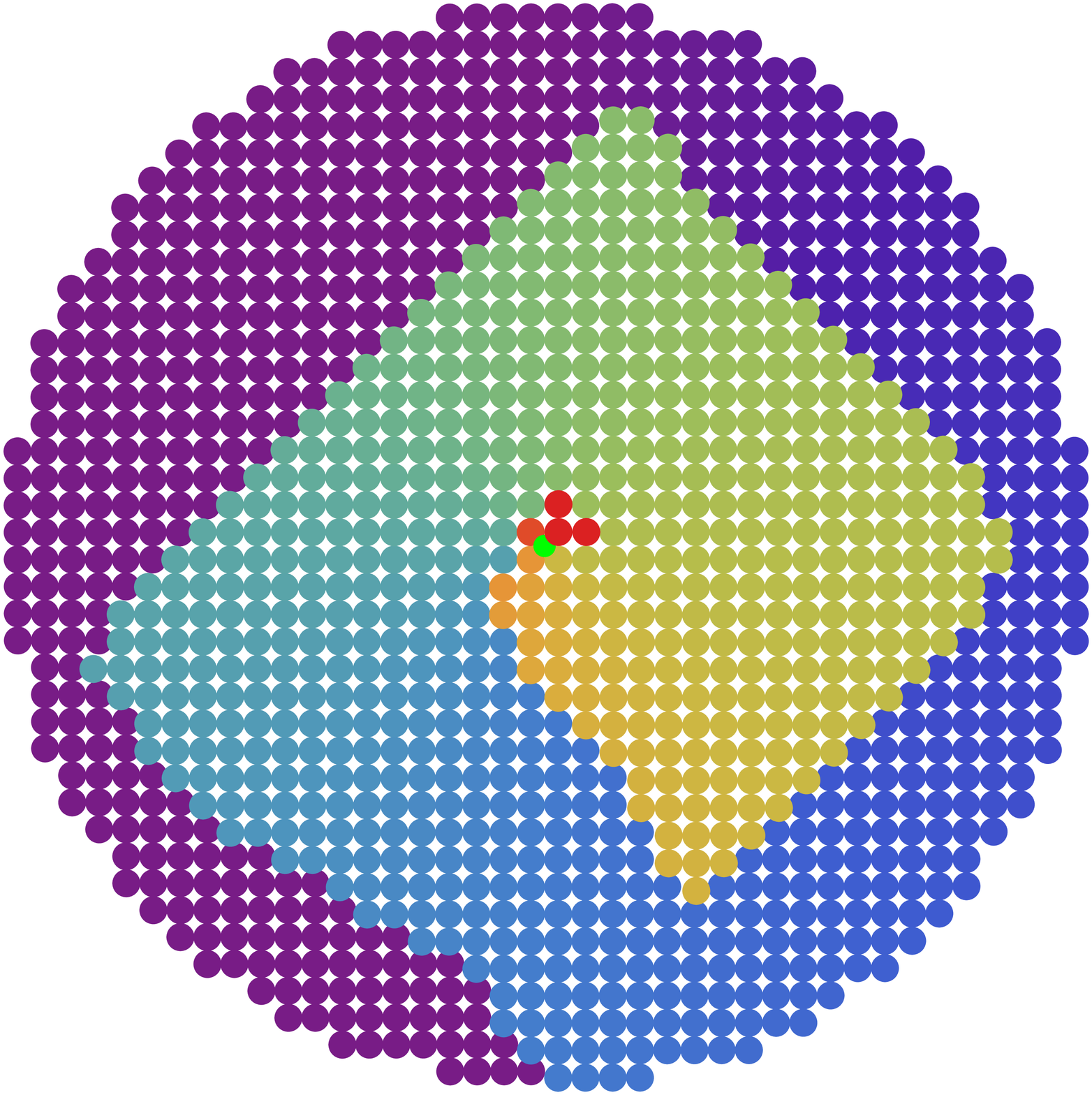}\\
    (c) \hspace{1.5in} (d)
    \caption{Growth around a screw dislocation starting from (a) initial film's profile, then after (b) 200 atoms, (c) 400 atoms, and (d) 600 atoms are deposited.}
    \label{fig:FirstFewRevolutions}
\end{figure}

The result from Method II, on the other hand, does not suffer from this setback. During its first revolution, whose development is depicted in Fig.~\ref{fig:FirstFewRevolutions} (b) and (c), the spiral ridge looks like a four-pointed star whose boundary is concave. These concave fronts are filled up quickly afterward into straight edges orienting along the preferred directions making the spiral rectangular in appearance. The growth fonts are exactly the same as those seen in the case of flat substrate growth. As time progresses, more and more steps are generated and the film looks like a rectangular pyramid. The distance between adjacent steps also decreases with time, as is typical seen in actual growth experiment~\cite{GerbAnseBednMannSchl91, SpriUetaFranBaue96} and also in the phase-field simulation~\cite{KarmPlap98}. With this particular boundary condition, we observe the stationary state when the width of successive steps is exactly one. Since the details of the growth evolution depend largely on the chosen diffusion rule (the modified WV in this case) and the boundary conditions, we shall postpone the full analyses of the dependence of film's growth on these choices for our future work.

It should be mentioned that the strange initial `side-branching' spiral has never been observed elsewhere, either in the phase-field modeling or in energetic-based kMC simulations of growth around a dislocation. We believe that this artifact is specific to our choice of diffusion rule. The shape, however, is similar to a two-dimensional kMC growth simulation around a nucleation site with atoms having a short average mean free path~\cite{XiaoIwanAlexRose91}. By employing a multiple registration scheme, the island boundary was smoothed out and the dendritic feature with four side branches could be seen. We believe that their striking resemblance albeit different physical processes may not be a coincidence but an implication about a connection between our probabilistic approach and the approach using short-distance diffusion with multiple registration scheme~\footnote{The latter was recently shown to be equivalent, and could provide an alternative approach to, simulating collective diffusion phenomenon on film growth~\cite{Reis10}.}. As a noise reduction technique, the scheme allows for a more probable site to be chosen since atom must visit a site repeatedly up to a certain number before it becomes a part of the film. Thus to some degree, the two approaches are similar. Further investigation is needed in order to quantify this connection.

\begin{figure}[htb]
	\centering
	\includegraphics[width=.47\textwidth]{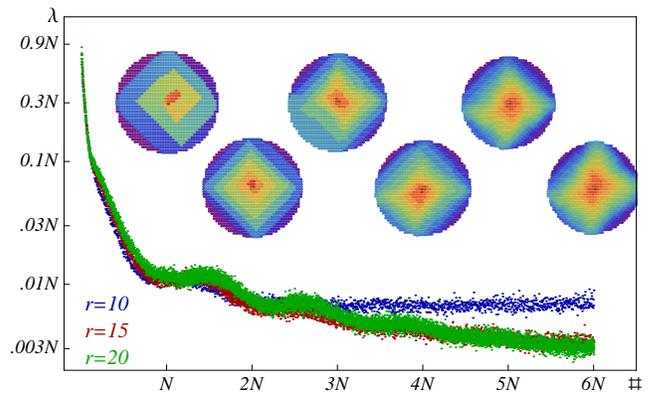}
	\caption{The average mean free path $\lambda$ simulated by Method I and measured in the units of the respected number of sites $N$ is plot as a function of subsequent drops up to the 6$N^\text{th}$ atom for circular substrate of radius 10, 15 and 20. The surface profile after every $N^\text{th}$ deposited atoms simulated by Method II are shown as points of reference.}
	\label{fig:NumOfSteps}
\end{figure}

As a demonstration of our probabilistic approach, Fig.~\ref{fig:NumOfSteps} shows the mean free path $\lambda$, or the average number of hops an atom makes until its incorporation to the most likely spot on the spiral ridge, computed through Method I. The simulations were performed on the substrate of radius $r = 10$, 15 and 20 atomic spacings bringing the total number of sites to $N = 316$, 716 and 1264 respectively. About $6N$ atoms were sequentially dropped in each case. Points on the graph are the results of the average over 800, 400 and 300 runs respectively. All three graphs are more or less on top of one another except for the tail of the $r=10$ case. Due to its small size, the system reaches its stationary state long before the other two cases. The very first atom has to hop on the order of $N \sim r^2$ before it reaches the dislocation ridge. The number of hops decreases very rapidly as the ridge starts to spiral. We notice a series of plateaus starting approximately at every $N^\text{th}$ atoms. A drop in the number of hops to the next plateau occurs as one or more of the ridge fronts are filled up and straightened out. Inside of a plateau region, new atoms take on average the same number of hops before reaching the spiral as its fronts propagate outward. This process continues until the stationary state is reached at which point atoms most likely would take at most a few hops before being incorporated into the spiral ridge. As points of reference, we include the shapes of the substrate, simulated by Method II, after $N^\text{th}$, $2N^\text{th}$, $\ldots$, $6N^\text{th}$ atoms are absorbed into the spiral.

A few remarks are in order before we end this section. In principle it is possible to obtain Fig.~\ref{fig:NumOfSteps} using the conventional approach by directly applying the WV diffusion rule to each step until an atom no longer moves, while recording its hop number. Doing so repeatedly in order to achieve the same statistics presented here, however, would be computationally very costly. We were able to produce the data used to create the above graphs using about four days of computer time on a single core processor. If one is not interested in carrying out any statistical quantities and only wants the evolution of the film's height profile, a much larger system can be simulated within a reasonable time. As mentioned earlier, in most cases the eventual resting site for each deposited atom may be obtained with minimal computation by simply looking at the structure of the underlying DAG.

\section{Conclusion}\label{S:Conclusion}
Based on Markovian hypothesis and Froebenius theorem, the limiting probability transition matrix for a random walker starting his trip with a given initial probability profile is obtained. We devised graph algorithms to automate the process so that it could be implemented on a computer. In the process, we discover an algorithm for finding the inverse of a certain class of stochastic matrices. Finally the formalism is applied to thin film's growth on a two-dimensional flat substrate and around a screw dislocation. The latter gives a usual spiral ridge with rectangular shape reflecting the underlying crystal structure in the limit where atoms are set to emerge at the most probable film's positions during growth. The result also suggests an interesting connection with the widely-used multiple registration technique in kMC simulations.






\appendix
\section{DAG and matrix inversion}\label{S:Inverse}
An algorithm such as that of Tarjan's which casts a matrix into the corresponding directed acyclic graph from where permutation matrix $\Qmat$ could be constructed, offers a new way of computing an inverse of a certain class of matrix. It is well known that the inverse of a triangular block matrix is given by
\begin{equation}\label{E:inverse}
    \begin{pmatrix}
        \Bmat_1 & \Bmat_2 \\ \Zero & \Bmat_3
    \end{pmatrix}^{-1} =
    \begin{pmatrix}
        \Bmat_1^{-1} & -\Bmat_1^{-1}\cdot\Bmat_2\cdot\Bmat_3^{-1} \\ \Zero & \Bmat_3^{-1}
    \end{pmatrix}.
\end{equation}
We shall use Eq.~(\ref{E:inverse}) as a basis for our analysis.

\begin{figure}[htb]
	\centering
	\includegraphics[width=.3\textwidth]{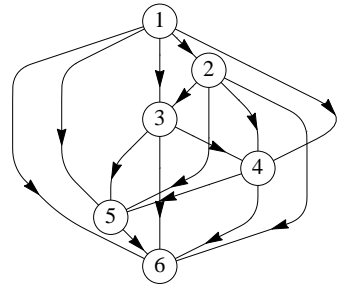}
	\caption{An example of a directed acyclic graph representing a matrix in its canonical form.}
	\label{fig:DAG_inverse}
\end{figure}

We are interested in finding the inverse of matrix $\Dmat \equiv \Id - \Amat$ where $\Amat$ is a sub-stochastic matrix whose summation of elements in each row is less than or equal to 1. We start by obtaining $\Qmat$ through Tarjan's algorithm. Matrix $\Qmat$ can be used to turn $\Amat$, through a simple change-of-basis, into 
\begin{equation*}
    \widetilde{\Amat} \equiv \Qmat^\top \cdot\Amat\cdot\Qmat = \begin{pmatrix} \Transient_1 & \Transient_{1,2} & \cdots & \Transient_{1,t} \\
							  \Zero & \Transient_2 & \cdots & \Transient_{2,t} \\
							 \vdots & \ddots & \ddots & \vdots \\
							 \Zero & \cdots & \Zero & \Transient_t \end{pmatrix}.
\end{equation*}
Matrix $\widetilde{\Amat}$ would have the same structure as, e.g., $\Tmattt$ shown in Eq.~(\ref{E:Tpermstructure}), where each $\Transient_i$ and $\Transient_{i,j}$ are irreducible. Fig.~\ref{fig:DAG_inverse} gives an example of the underlying DAG of $\widetilde{\Amat}$ for $t=6$, assuming that all of the upper triangular block matrices are non-zero.

Let $\Mmat_i \equiv (\Id-\Transient_i)^{-1}$ and $S(i,j)$ be the set of all possible subsets of $\{i,i+1,\ldots, j\}$ with $i$ and $j$ as the first and the last elements, and is listed in increasing order. For example, $S(1,4) = \{\{1,2,3,4\},\{1,2,4\},\{1,3,4\},\{1,4\}\}$. One can recursively apply Eq.~(\ref{E:inverse}) to compute the inverse of
\begin{equation}\label{E:CanonMat}
    \widetilde{\Dmat} = \Id - \widetilde{\Amat} = \begin{pmatrix} \Id\! - \!\Transient_1 & -\Transient_{1,2} & \cdots & -\Transient_{1,t} \\
							  \Zero & \Id\! - \!\Transient_2 & \cdots & -\Transient_{2,t} \\
							 \vdots & \ddots & \ddots & \vdots \\
							 \Zero & \cdots & \Zero & \Id\! - \!\Transient_t \end{pmatrix}.
\end{equation}
It is straightforward to show that the $(i,j)^\text{th}$ block component of $\widetilde{\Dmat}^{-1}$ is given by
\begin{equation}\label{E:Dinv}
    \big[\widetilde{\Dmat}^{-1} \big]_{i,j} = \begin{cases}
         \Mmat_i & i = j, \\
         \Mmat_i\cdot  \!\! \displaystyle \sum_{s\in S(i,j)} \!\! \prod_{k=1}^{\parallel s \parallel - 1} \!\! \left(\Transient_{s_k,s_{k+1}}\cdot\Mmat_{s_{k+1}} \right) & i < j, \\
         \Zero & i > j.
     \end{cases}
\end{equation}
The sum in the case of $i<j$ is taken over all members $s$ of $S(i,j)$, where $s_k$ denotes the $k^\text{th}$ element of $s$ and $\parallel s \parallel$ denotes the number of elements. Once the inverses of all block elements are computed and the whole $\widetilde{\Dmat}^{-1}$ is assembled, one can simply perform another change of basis to shuffle all elements back to their original orders.

Eq.~(\ref{E:Dinv}) may be conveniently read off from the structure of the DAG of $\widetilde{\Amat}$. To obtain the $(i,j)^\text{th}$ block component, one simply traverses the graph from node $i$ to $j$ through all possible routes. Each visit to node $k$ corresponds with $\Mmat_k$. Each passage through an edge from $l$ to $m$ corresponds with $\Transient_{l,m}$. The final result is the sum over these routes. This method amounts to graph traversal which is a common routine in graph programming.

Our method of finding an inverse not only is simple, but also reveals the fundamental structure of the matrix. Moreover, in some problems, only a small subset of inverse matrix elements are needed. Our method would tremendously reduce the amount of computation because only a few $\Mmat_i$'s may be required. Finally we should point out that directly finding $\Amat^{-1}$ is not much harder than finding $\Dmat^{-1}$. The additional difficulty arises in keeping track of extra minus signs that crops up depending on whether the number of $\Transient_{l,m}$'s in an expression is odd or even. This could be done simply by counting the number of nodes visited during the traversal.

\begin{acknowledgments}
The author is grateful to Assoc.~Prof.~Dr.~Udomsilp Pinsook for his insightful input and valuable suggestions. Fundings from Thailand Center of Excellence in Physics and the Thailand Research Fund Grant No. MRG5580245 are acknowledged.
\end{acknowledgments}

%

\end{document}